\documentclass[sigconf]{acmart}
\AtBeginDocument{%
  \providecommand\BibTeX{{%
    \normalfont B\kern-0.5em{\scshape i\kern-0.25em b}\kern-0.8em\TeX}}}

\copyrightyear{2023}
\acmYear{2023}
\setcopyright{rightsretained}
\acmConference[CHI EA '23]{Extended Abstracts of the 2023 CHI Conference on Human Factors in Computing Systems}{April 23--28, 2023}{Hamburg, Germany}
\acmBooktitle{Extended Abstracts of the 2023 CHI Conference on Human Factors in Computing Systems (CHI EA '23), April 23--28, 2023, Hamburg, Germany}\acmDOI{10.1145/3544549.3582741}
\acmISBN{978-1-4503-9422-2/23/04}

\usepackage{tfrupee}
\usepackage{sectionbreak}

\begin{document}

\title[Should Computers Be Easy To Use?]{Should Computers Be Easy To Use? Questioning the Doctrine of Simplicity in User Interface Design}

\author{Advait Sarkar}
\affiliation{%
  \institution{University of Cambridge, University College London, and Microsoft Research}
  \country{United Kingdom}
}


\begin{abstract}
That computers should be easy to learn and use is a rarely-questioned tenet of user interface design. But what do we gain from prioritising usability and learnability, and what do we lose? I explore how simplicity is not an inevitable truth of user interface design, but rather contingent on a series of events in the evolution of software. Not only does a rigid adherence to this doctrine place an artificial ceiling on the power and flexibility of software, but it is also culturally relative, privileging certain information cultures over others. I propose that for feature-rich software, negotiated complexity is a better target than simplicity, and we must revisit the ill-regarded relationship between learning, documentation, and software.
\end{abstract}





\maketitle

\section{The doctrine of simplicity}


To give the doctrine of simplicity a single definition is difficult, as it appears in many variants and guises. It encapsulates a certain ideology and set of values about computer software. It takes the form: ``computers should be easy/natural/simple to use/learn''. A critical reader will seize upon the many possible combinations of words in the preceding sentence, and point out that each can be interpreted differently. ``\emph{natural} to \emph{use}'' is not the same as ``\emph{easy} to \emph{learn}''; certainly the design of a user study to test either proposition will radically differ. I acknowledge these distinctions and request the reader to indulge this temporary generalisation, because I hope to make clear that the specific variant of the doctrine is not critical to the following discussion, rather it is the ethos that this set of statements encompasses, their origin, and their implications, that I wish to bring into focus.

A corollary to the doctrine of simplicity is the doctrine of gradualism, that computer users can be led to difficult concepts via easier stepping stones. As a design principle, this translates as: at each incremental stage of the user learning journey, give the user the smallest necessary amount of learning, information, flexibility, and power. This is ostensibly inspired by the way in which concepts are presented in a particular order in a textbook or a school curriculum, although the idea of ``conceptual prerequisites'' \cite{griffin1994rightstart}, that some concepts inherently `precede' others, is a contested pedagogical principle. As implicit (and sometimes explicit) pedagogues, software and its designers manifest a particular philosophy of learning and teaching.


A reader who has worked for any period of time in human-computer interaction research or design will have encountered the doctrines of simplicity and gradualism. Software design can be even viewed as the search for metaphors to make hard concepts graspable \cite{blackwell2006reification}. Similar principles and values are articulated formally and informally in design theory, design research, and in practice. Let me draw upon a few examples to demonstrate the point that simplicity and gradualism have been highly influential ideas in human-computer interaction research.



Early explainable AI research discovered a limit to the amount of information in an explanation before the user became overwhelmed \cite{kulesza2013too, sarkar2022explainable}. This research formed the partial basis for guidelines for human-AI interaction design \cite{amershi2019guidelines}. Simplicity is achieved by limiting the quantity of information presented, or by limiting (or altering) its quality. Approaches to information limiting in explainable AI (e.g., ``don't overwhelm'' \cite{kulesza2013too}) and in data visualisation (e.g., ``overview first, zoom and filter, then details-on-demand'' \cite{shneiderman2003eyes}) are arguably more focused on quantity, whereas some interfaces apply qualitative gradualism, using the theory of multiple representations \cite{ainsworth1999functions}, to guide learners through increasing levels of representational abstraction \cite{stead2014learning,hermans2020hedy,sarkar2018calculation}.

In end-user programming and programming education, there is the concept of an `abstraction gradient' \cite{green1996usability}, along which programming language features can be arranged from low to high abstraction. Various projects improve the learnability of programming systems by beginning the abstraction gradient at a low level \cite{sarkar2016phd}, systematically introducing syntactic abstractions \cite{hermans2020hedy, miller2016gradual}, or using multiple representations at different levels of abstraction (an approach rooted in mathematics education) \cite{stead2014learning, bau2015pencil, sarkar2018calculation, ainsworth1999functions}. Petricek characterises ``no-code'' or ``low-code'' programming systems, such as HyperCard, by the ``substrates'' they offer to change the software; in general, the larger the scope for change, the higher the difficulty of operation \cite{petricek_2022}. Successful systems, he argues, apply gradualism: \emph{``[...] the visual environment is structured in terms of several user levels [...] that gradually unlock more advanced features of the system. [...] This somewhat reduces the gap [...] that you need to bridge if you need more complex change, because you are staying in the same substrate [...] you do not get overwhelmed by all that is available, as more advanced features are initially hidden''}.

As personal computing grew in the late 1980s and early '90s, coinciding with the widespread deployment of graphical user interfaces, the design of documentation and software help systems became a focus for HCI research. Key findings from this research would shape attitudes towards software learnability for decades to come. Carroll and Rosson's pioneering work identified the ``paradox of the active user'' \cite{carroll1987paradox}, who fails to make progress because they prefer to tinker with the software rather than read documentation. Rettig proclaimed that \emph{Nobody Reads Documentation} \cite{rettig1991nobody}. Out of these insights developed the paradigm of ``minimalist instruction'' \cite{carroll1990nurnberg} and many ingenious innovations in graphical user interface design that are taken for granted today, such as the tooltip \cite{farkas1993role}.

Unfortunately, as we will see in the following sections, these successes also instilled a kind of designer-chauvinist idea that great design can all but eliminate the need for learning. In \emph{Training Wheels in a User Interface} \cite{carroll1984training}, a gradualist proposal that complex features can be gated until users have achieved a certain level of expertise, Carroll and Carrithers demur about its applicability as a permanent solution, stating: \emph{``a more significant and longer term goal in system design is that of defining a user interface architecture with which to confront learnability and training issues, and with which to eliminate the need to retrofit for learnability in the first place''}.

With the growth of touchscreens, cameras, sensors, and multi-modal computing, it became possible to control computers through gestures. Based on the principle that gestures should require as little learning as possible, in the mid-2000s the field of gesture elicitation studies emerged \cite{wobbrock2005maximizing}, spawning over 200 studies \cite{villarreal2020systematic} aiming to discover gestures and symbols so natural that they pre-exist in the user's repertoire before they encounter a system at all.

Here I have highlighted a few manifestations of the doctrines of simplicity and gradualism that I have encountered in my own research. I believe that many examples besides these can be found in many corners of human-computer interaction, user experience, and design practice. They are deeply ingrained in our research practices and widely-used instruments including the System Usability Scale \cite{brooke1996sus} (cited over 14,000 times) which includes questions such as \emph{``I thought the system was easy to use''} and \emph{``I would imagine that most people would learn to use this system very quickly''}, and the NASA TLX cognitive load index \cite{hart1988development,hart2006nasa} (multiple publications, together cited over 18,000 times), including questions such as \emph{``How much mental and perceptual activity was required (e.g., thinking, deciding, calculating, remembering, looking, searching, etc.)? Was the task easy or demanding, simple or complex, exacting or forgiving?''} and \emph{``How hard did you have to work (mentally and physically) to accomplish your level of performance?''} Many researchers, myself included, rely heavily on these instruments to evaluate designs. But is it always an indicator of superior design, when someone can learn to use a system very quickly? Is it always a better interface where the task is easy, simple, and forgiving, and requires little ``mental and perceptual activity''?

These ideas are almost axiomatic, and seem like unquestionable tenets of good user interface design. Yet this particular set of values can be traced back to specific, contingent events in the cultural and commercial history of computing.

\section{How did we get here?}
\label{sec:history}

\begin{figure*}[t]
  \centering
  \includegraphics[width=\linewidth]{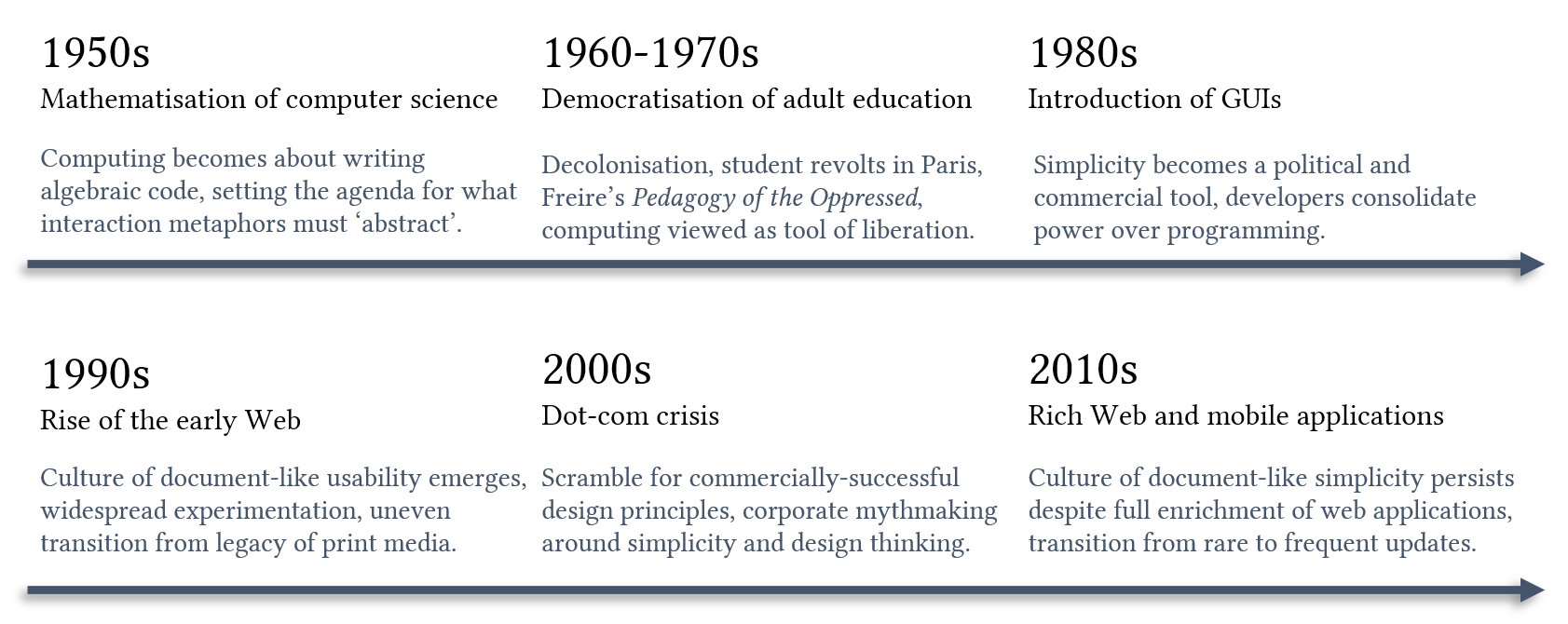}
  \vspace{-0.8cm}
  \caption{
    An overview of developments in the history of the doctrine of simplicity.
  }
  \label{fig:timeline}
  \Description{A timeline diagram with arrows running from left to right depicting time. On the timeline are marked examples of events from each decade from the 1960s to 2000s.}
  \vspace{-0.5cm}
\end{figure*}

So far we have briefly encountered the doctrines of simplicity and gradualism as we know them today, and hopefully the examples above have shown their pervasive influence in HCI research and practice. The title of this paper indicates that we will be questioning that influence. As with any proposal for reform, we risk committing the fallacy of removing Chesterton's proverbial fence \cite{chesterton1929thing}, proposing to dissolve an apparently outmoded institution, only to discover the complex systemic interactions that necessitated it. Thus it is worth attempting to understand why and how we got here in the first place. A rough outline of the discussion that follows is presented in Figure~\ref{fig:timeline}.

The first important development in the history of user interface design was the conceptualization of the textual (algebraic-sentential) programming language as the basis for instructing computers. In conceiving of programming as writing code, early computer scientists implicitly set forth the agenda for interface designers as code being the underlying complexity that design metaphors ought to abstract away. But this was scarcely inevitable: Arawjo traces key cultural, commercial and technical constraints, including institutional directives, the requirements to retrofit punch card machines and typewriters, and even aesthetic xenophobia (Frege's two-dimensional notation was ridiculed as ``Japanese'' \cite{mendelsohn2005philosophy}), which led to the dominance of the sentential paradigm over visual, diagrammatic alternatives \cite{arawjo2020write}.

Moreover, within textual programming languages the view of programming as operationalised mathematics is only one possibility. Babbage mostly conceived of programming this way, though Lovelace saw further, speculating in a celebrated example that programs could be used to manipulate music \cite{snyder2012philosophical}. Nonetheless, in the 1950s the nascent computer ``science'' was proximal to many potential disciplinary homes, from electrical engineering, physics, astronomy, and biology to communications, espionage, archival, and art. The mathematisation of computer science (and consequently its principal working material, the programming language) was a result of computer science's struggle for disciplinary legitimacy \cite{ensmenger2012computer}. Eager to be taken seriously as a science, and eager to capture investment from government agencies and private enterprise, the discipline aligned itself to the comforting objectivity and inarguable utility of applied mathematics. The discipline was gradually so entrenched in this particular episteme \cite{Foucault1966-ey} that it became virtually impossible to imagine any alternative.





Yet it was in the imagination of an alternative that interface design emerged. The practice of programming computers began as complex, requiring manuals and study. The use of language to hold power and separate classes is a tale that recurs throughout human history. The Egyptian priestly class kept power through guarded knowledge of the sacred hieroglyphs, hindu brahmin priests kept power through their rites and incantations \cite{mosca1939ruling}, catholic priests kept power over the interpretation of the Bible through their knowledge of ecclesiastical Latin \cite{jensen1996humanist}, and in modern society, a new class of scientific priests maintains power through jargon and institutional gatekeeping \cite{foucault1969archeologie}. The emancipatory era of the 1960s and '70s was characterised by the rapid withdrawal of global colonial powers from physical occupation \cite{duara2004introduction}, tumultuous developments in attitudes to student empowerment in adult education \cite{vetter2019global,freire1968pedagogia}, and the increasing export of American democratic values to solidify the Western postwar hegemony \cite{nye1990soft}. Swept along amidst these vast social and political currents, personal computing symbolised the new information age, and it became apparent that a democratic future was one in which everyone could use a computer, and that meant democratising the language of instruction.


It was not just the starry-eyed ideal of liberal democracy that coaxed computing out of the hands of scientist-priests, but also the prospect of fabulous wealth. Bill Gates took on the mission to place ``a computer on every desk'' \cite{gates2007robot}. Though Xerox did not rise to the opportunity for commercialising Alan Kay's visionary research at PARC \cite{myers1998brief}, the introduction of the graphical user interface in the 1980s by Apple and Microsoft became the next formative step in the doctrine of simplicity. Simplicity and its perception became a marketing and competitive advantage. Apple's marketing materials focused on the simplicity of the operating system, lampooned competitor systems as requiring stacks of heavy manuals,\footnote{For example: \url{https://www.youtube.com/watch?v=3vq9p00T08I}} later adopting the taglines ``there is no step 3'' and ``it just works''. Apple's design interpreted simplicity as radical minimalism in software and hardware (often at the expense of actual usability, as noted in the case of the single-button mouse, and the buttonless iPod shuffle \cite{price_2022}). 
The adoption of simplicity as a design principle in this era invoked, sometimes explicitly, a notion of political superiority. Apple's famous ``1984'' ad compared the usability of the Macintosh to the revolution in an Orwellian communist dystopia (with IBM, the incumbent giant, in the position of the repressive state). It has been widely discussed that Steve Jobs' minimalism, which spanned his attire, his communication style, and his products, was heavily influenced by his personal experiences with ascetic Buddhism \cite{robinson2015appletopia}.


But more important than these explicit politicisations were the effects that the simplicity shift had on society, research and practice. Winner sets out two mechanisms by which technology exerts political influence \cite{winner1980artifacts}, which we can see in the case of the graphical user interface. First, the specific design features of a technology or device enable certain forms of power (low bridges permit only certain vehicles to pass underneath). Second, the decision to adopt a technology or not, for instance a certain type of machine, or a certain standard, lends itself to certain forms of power being invoked because that is what is needed to make the technology work (a ship's captain requires absolute authority to manage the control of the ship in crisis). In the case of the graphical user interface, its commercial success rapidly eliminated other forms of interface. At the same time, the power of programmers and the software industry was consolidated, because while graphical user interfaces were usable, they were fundamentally restricted; end-users could not describe new behaviours -- you need a programmer for that. End-user programming research has ever since battled this limitation \cite{ko2011state}, with the development of powerful paradigms such as programming-by-demonstration \cite{kurlander1993watch} and programming-by-example \cite{lieberman2001your}, and a few widespread commercial successes, such as the spreadsheet. Yet the legacy of mathematisation and algebraic-sentential programming still looms over these efforts; it is still that particular form and vision of computing that we seek to make more accessible.



Desite Apple's mockery of manuals, through the 1980s and '90s most software still came with manuals, and it was culturally accepted that some explicit training and learning investment was required to use computer software. This was reasonable given that a lot of software was used in a business context, and the longevity of software was high. Updates were rare and expensive. Due to the distribution bottleneck of floppy disks, CD-ROMs, and the retail supply chain, software packages were updated once a year at most. Software was largely sold with a permanent license, rather than the now-predominant subscription model. In such a climate, it made sense to approach software as a tool with relatively fixed behaviour, because time invested in learning had a high likelihood of a long-term return in productivity.

This began to change in the late 1990s with the World-Wide-Web. Initially conceived as a method for sending communications and documents, websites were simple and usable by default --- as usable as a document --- to use the Web, one needed only to be able to read and write. Yet unlike static documents, the Web did have minimal levels of interaction: links could be clicked, forms could be submitted. From this humble basis emerged a tidal wave of speculative applications, from search engines, to news media, to e-commerce. This Wild-Wild-West came to an abrupt halt with the burst of the dot-com bubble in 2000, when it became apparent that not just any Internet business would succeed.

A business that had succeeded though, was Google. In the aftermath of the dot-com bubble a scramble began to discover the formula for success on the Web. Google's success, in part, began to be attributed to its minimalist design, which stood in stark contrast to the ``cluttered'' interfaces of its competitors: Yahoo, AltaVista, Lycos, and Ask Jeeves, which inherited their design from book indexes and print media. This began a period of intense mythmaking in design culture where simplicity became conflated with usability, and was exalted as the key to building commercially successful software. This was the era of Jenson's \emph{The Simplicity Shift} \cite{jenson2002simplicity} and Maeda's \emph{Laws of Simplicity} \cite{maeda2006laws}. The influential design firm IDEO began its crusade to instill ``design thinking'' in every corporation, having expanded its interests from designing chairs and toothbrushes to management consultancy \cite{nussbaum_2004}. Products and their designers began to assume the mantle of instruction manuals and teachers, and a product that needed effort to learn was not only considered a commercially inferior prospect, but also reflective of poor design.

The early Web had trained users to expect document-like levels of simplicity, but in the 2000s, with Flash, Java Applets, and JavaScript, the interactive potential of Web technologies grew. Social media sites such as Facebook and Twitter were the great winners of this new Web ecosystem and it is no coincidence that they remain key players in advancing infrastructure for enriching web apps, such as React and Bootstrap. Similarly Google, which did not own a desktop operating system, poured resources into building Chrome, which subsequently became the dominant Web browser. Google and other browser vendors, through their involvement in the W3C consortium and WHATWG groups, pursued key capabilities in multimedia, local storage, multi-threading, and high-performance JavaScript, facilitating the functional parity of Web applications with desktop applications. 


Web ``sites'' to be visited and Web ``pages'' to be read became Web ``applications'' to be used \cite{laurel2013computers}. But while these applications are as rich, capable, and complex as ``native'' desktop software, the cultural expectations of the static Web persist. When users open an application in a browser, they expect to use it without investing effort to learn the interface, as though it were a document. Through gradual enrichment, the designers of web apps became trapped, like the proverbial frogs in boiling water, between their ambitions for their software and the expectations of users.

Rich Web applications transformed the status quo of rare updates and software continuity; the Web enabled instant and continuous deployment of new features. Mobile operating systems were born into this new regime and adopted the same model. Through the 2010s, traditional desktop software also began adopting the continuous deployment engineering model (and its attendant economic model of subscriptions), facilitated by the rapid growth of broadband and 3G/4G mobile data networks. With the benefits of frequent updates came another cultural shift in the value proposition of investing time and effort into learning software. What if a feature's appearance or behaviour changes? What if it is removed? With software subject to constant change, it becomes difficult to predict whether effort invested in learning will pay off. This puts further pressure on software to be simple and self-evident.

The research focus of HCI (and CHI in particular) had aligned itself with this landscape of discretionary commercial software use, thus separating it from its predecessor, human factors, which largely targeted the design of systems used within organizations by employees with little choice in the matter \cite{grudin2009ai}. Weiser's research agenda of ubiquitous computing \cite{weiser1991computer}, a vision of ``quiet'', ``invisible'', and ``seamless'' interfaces, became entrenched as a design objective. SIGDOC, the ACM special interest group on documentation, home to much early research on software help systems, was threatened with obsolescence and rebranded itself as ``design of communication'' \cite{mehlenbacher2011evolution}. This accompanied a broader turn in computer science education away from direct instruction towards constructivist learning, the idea that computing is best learned through experimentation, not instruction \cite{hermans2018explicit}, which is reflected in contemporary studies of software skill acquisition \cite{sarkar2016constructivist,sarkar2018spreadsheetlearning,sarkar2020spreadsheet}. The practical nature of academic HCI also played a role: longitudinal studies are hard, single-session lab studies are (relatively!) easy. Short studies are more conducive to the annual drumbeat of conference submission cycles. We develop too many new systems for each to be evaluated in long-term use; the ability to rapidly test and explore various design spaces is an asset to our discipline, but it places pressure on design to make its value evident on the basis of brief and shallow encounters.







\begin{figure}[t]
  \centering
  \includegraphics[width=\linewidth]{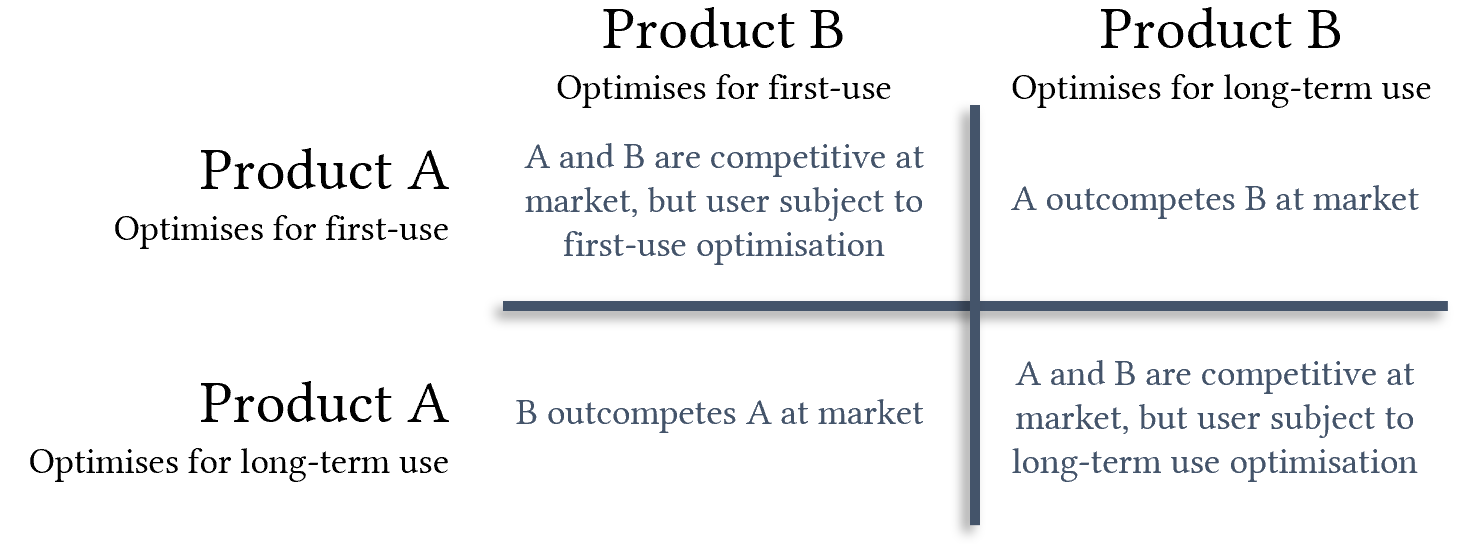}
  \caption{
    Product design in a commercial environment that rewards simplicity can result in a poor equilibrium. The Nash equilibrium (top left) is not the Pareto efficient state (bottom right).
  }
  \label{fig:nash}
  \Description{A game theory diagram showing...}
\end{figure}

In this current climate, where users evaluate potential software purchases and subscriptions within minutes, and impressions of websites are made in less than a second \cite{lindgaard2006attention}, first impressions of usability matter and are critical to commercial success. But in certain cases, this creates a moral hazard to make software appear \emph{more usable} by making it \emph{less useful}. The design of commercial software aimed at discretionary users has fallen into a Nash equilibrium which is not Pareto efficient, a situation which most introductions to game theory illustrate using the example of the prisoner's dilemma \cite{osborne2004introduction}. Each prisoner can unilaterally maximise their outcome by confessing, but this results in a sub-optimal equilibrium, where the optimal solution would have been for both to cooperate. As in the prisoner's dilemma, products optimised for long-term use can lose out to those which appear easier to use on a preliminary evaluation (Figure~\ref{fig:nash}). This market environment puts pressure on designers to hyper-optimise for making products easy to use without learning investment, but which may not result in the best long-term user experience or empowerment.


\section{What we lose when software is `easy'}

In some cases the benefits to users of a competitive market pressure to make software `easy' are obvious; users clearly stand to gain when a brilliant metaphor or interface technique makes the same capabilities available to the user with lower cognitive and learning costs. But what might they lose when designers adhere to these doctrines?

What about software whose entire purpose is to create radical leaps in users' thinking? What if such software requires users to shift their worldview in order to use it? This might bring to mind sweeping concepts such as Kuhn's paradigm shifts \cite{kuhn1970structure}, for instance the transition away from phlogiston theory in the 1770s with Lavoisier's isolation of oxygen. Or Galileo's shifts in natural interpretations, for instance the re-evaluation of sense-perception as a source of truth regarding motion \cite{feyerabend2010against}. Software may even aim to serve as Wittgenstein's ladder \cite{sep-wittgenstein}, which a learner must climb only to throw away, achieving an enduring cognitive extension \cite{hernandez2019ai}.



But we needn't reach for such extremes: there are many common examples of concepts, skills, and tools that are hard to learn, which may take years, and thousands of hours of deliberate practice \cite{anders2008deliberate}, but those who learn them have transformed themselves and their minds. Mathematics, language, writing, playing a musical instrument, playing a sport: almost any human activity worth doing falls into this category.

It might seem like a stretch to draw an analogy between playing a musical instrument and using software, but in fact some kinds of software also fall into this category. This includes spreadsheet software, such as Microsoft Excel and Google Sheets, design software such as Adobe Photoshop and Autodesk AutoCAD, digital audio workstations such as Apple's Logic Pro and Ableton Live. It includes specialist applications in domains as wide as accounting, urban planning, architecture, astrophysics; indeed almost every discipline and profession is served by some unique and distinctive software tools that form a core and indispensable part of its practice. These applications are used intensively, and over long periods of time, often entire careers. Their feature set is vast and complex, and users need to work to build skills. Often, professional identity and status is deeply entwined with the skillful use of specific software, resulting in a fascinating power dynamic between users, software, and its designers \cite{nouwens2018application, sarkar2022end}. Users develop craft practices, idioms, and pattern languages for using such software. Such software often has thriving online and offline communities where expert knowledge is exchanged. While individual practitioners may make considered software purchase decisions based on long-term evaluation, even software of this kind is subject to commercial pressures, as companies seek to expand market share by capturing casual usage, and increasing the appeal of software in B2B sales settings, where software procurement decisions are often made by managerial non-practitioners on behalf of the entire organisation.

As a shorthand, let us call this particular kind of software \emph{praxisware}. The term ``praxis'' is widely used, and often loosely used simply to mean ``practice''. I am drawing particularly from the notions of praxis due firstly to Sartre \cite{Soper1986-dy}, who situates praxis within division of labour and for whom praxis constitutes activities that consolidate and protect class and profession, and secondly to Freire \cite{freire1968pedagogia}, who situates praxis within education and empowerment and for whom praxis constitutes activities of ``reflection and action directed at [societal] structures to be transformed''.


It is in praxisware that the tradeoffs between superficial simplicity and long-term user experience are most acute. This software aims to facilitate rich and complex activities that are antagonised by the ideal of simplicity. As Stolterman notes~\cite{stolterman2008nature}: \emph{``Humans seem to seek and enjoy certain experiences of complexity. In some contexts, complexity may be understood as richness, generally found to be a positive and desired quality. [...] The simpler an environment is, the easier it is to understand and deal with, but at the same time, the more it lacks the richness and stimulus that we seem to appreciate and enjoy''}. Norman's foreword to Laurel's \emph{Computers as Theatre} explains: \cite{laurel2013computers}: \emph{``[...] we get the greatest pleasure from our ability to overcome early failures and adversaries. If everything runs perfectly and smoothly with no opportunity to deploy our powers and skills, pleasure is diminished. Human emotion is sensitive to change; starting low and ending high is a far better experience than one that is always high. Is this a cry for deliberate placement of obstacles and confusions? Obviously not, but it is a cry for a look at the temporal dimensions, at engagement, agency, and the rise and fall of dramatic tension.''}


This critique of simplicity echoes previous critiques voiced in third-wave HCI, which questioned values of `seamlessness' and Weiser's idea that technology should be quiet and fade into the background. In reaction it was pointed out that in some scenarios, `seamful' design might be preferable, to take advantage of the fact that user activities often interweave heterogenous media \cite{chalmers2004seamful}. Ambiguity was not universally to be avoided, but could be drawn upon as a design resource to create engaging and thought-provoking experiences, and practically address technological limitations \cite{gaver2003ambiguity}. More recently, with growing concerns around dispossession and loss of agency to AI algorithms, researchers have proposed that designs could be deliberately antagonistic as a societal intervention \cite{hollanek2019non}: \emph{``Could `non-user-friendly' design successfully harness the feeling of confusion and dissatisfaction to raise political awareness, to cause a cognitive glitch?''} Eric Li of the Museum of Modern Art, drawing upon Guy Debord's concept of deriv\'e (drifting), conjectures \cite{li_2021}: \emph{``But what if we [used] design to provoke a type of thinking that art has provoked for centuries? [...] Imagine that, when you go to your browser, the furniture and interface elements that greet you aren’t there to speed you onto your next destination on the Web. [...] What if it were filled with a random work of art? Time slows down [and you] pause, enjoying the sights, before diving back into the ever-flowing river of information. [...] Perhaps, like a d\'erive, not the most direct route, but far more enjoyable. A breath of air.''}

Computer games are another genre of software that are often complex, and have steep learning curves, but are nonetheless tolerated (even enjoyed) by users and are commercially successful. Researchers have attempted to understand the efficacy of game tutorials \cite{andersen2012impact}, and proposed that game heuristics such as challenge, fantasy, and curiosity might be applied to software design more broadly \cite{malone1982heuristics}. Yet there remains a fundamental difference: games are \emph{experiential} software, interacted with purely for their own sake, where users are intrinsically motivated to learn. In contrast, productivity software is used as part of a process that represents and manipulates outcomes external to the software itself, i.e. in the ``real'' world. As Laurel puts it, \emph{````Productivity'' as a class of applications is better characterized, not by the concreteness of outcomes, but by their seriousness vis-à-vis the real world''} \cite{laurel2013computers}. For this reason, games do not generally fall into the category of praxisware, and are not subject to the same historical and commercial influences or cultural expectations. This does not mean that there is nothing to learn from game design, but it does mean that we cannot expect that importing strategies for managing software complexity from games will be straightforward, or effective.

\centerline{\rule{0.3\linewidth}{.2pt}}
What is `easier' or `harder' is often a matter of perspective shift that only comes with expertise of a new paradigm. The enlightenment was partly fuelled by algebraic mathematical notation that replaced the earlier form of writing mathematical statements using long-form naturalistic language sentences \cite{okrent2009land, snyder2012philosophical}. With the benefit of hindsight, algebraic notation clearly made mathematics easier: symbolic manipulation unlocked efficiencies and scaffolded thinking in a manner that was simply impossible with the old notation. But using ordinary language to describe mathematics is also in some sense easier, because there is no new notation and its attendant manipulations to learn; this is precisely why it had been used for so long in the first place. Thus a tool that is initially easier (natural language sentences) may have a glass ceiling, and a tool that is initially harder (algebraic notation) can, after a perspective shift, make subsequent tasks easier.



Even to describe language or writing as ``easy'' or ``intuitive'' ignores the fact that learning to speak and write a language takes years of practice, something that is apparent to anyone who has tried to learn a second language after childhood. Chomskians might argue that the capacity for language is innate, but they will not argue the same for the ability to speak any \emph{particular} language. In text entry research, a language-adjacent corner of human-computer interaction, the importance of evaluating not the initial usability of a system but rather its emergent usability over time and learning investment is well-documented. Researchers reference the ``power law'' of learning (or practice) \cite{newell1981mechanisms} which describes how a user's proficiency improves over time, with practice and learning. Ward et al. propose that the slope of this curve, rather than the intercept, is a superior indicator of a text entry system's quality \cite{ward2000dasher}. At the same time, it is acknowledged that the commercial acceptance of such systems hinges on first-use and early-use experience (i.e., the intercept) \cite{mackenzie1999text}. Text entry is a microcosm that reflects the simplicity trap.



O'Hara et al. critique the rhetoric of ``naturalness'' that pervades research in touchless user interfaces \cite{ohara2013naturalness}, finding that rather than being a \emph{``representation of our gesture''} and the \emph{``ability to infer intent''} to facilitate the \emph{``exchange of information''}, naturalness arises from \emph{``the practices of specific communities in particular social settings [...] we need to approach the design of these systems in terms of how they might allow a beneficial reconfiguration of practices and how we experience the world in new ways accordingly.''}\footnote{Their critique is somewhat more sophisticated than mine, which contends that claims of ``natural'' interfaces should consider how natural it can be to coax a shard of electrified silicon to do your bidding.} In the case of praxisware, notions of objective simplicity and information exchange are antagonistic to the community-based and often professionalised nature of its use.









\centerline{\rule{0.3\linewidth}{.2pt}}
The ideal of simplicity as a goal for software design is culturally relative. Simplicity is interpreted and enacted at several levels of software design, from aesthetic, to functional, to the conceptualisation and unitisation of the user experience across a computing platform.

\begin{figure*}[t]
  \centering
  \includegraphics[width=\linewidth]{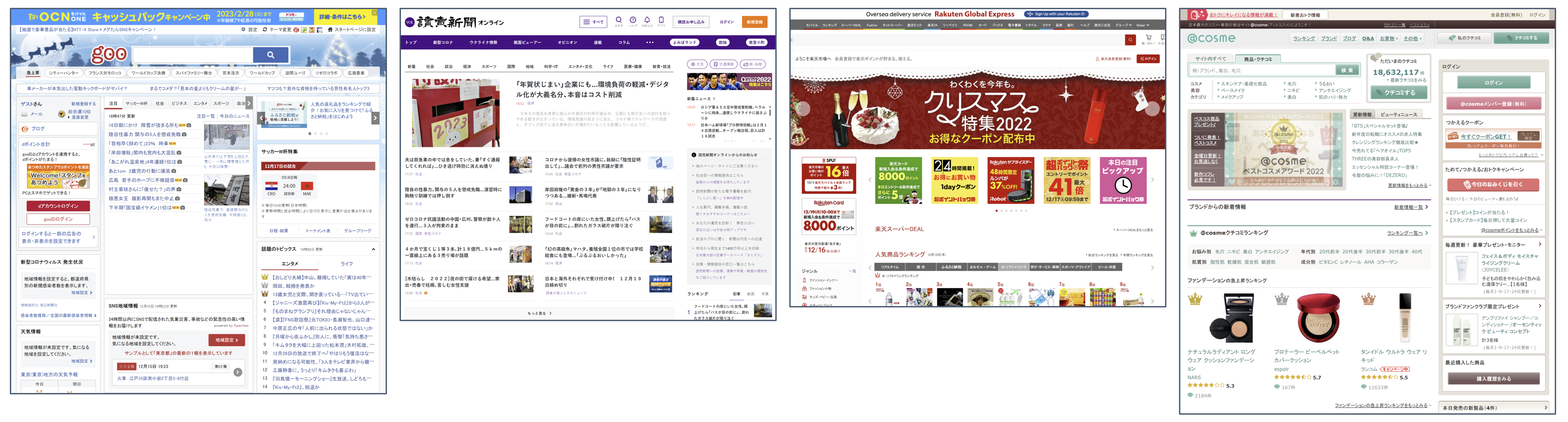}
  \caption{
    A sample of screenshots of popular Japanese websites, an alternative manifestation of simplicity (December 2022). Sources (left to right): \url{https://www.goo.ne.jp/} \copyright~NTT Resonant Inc.,
    \url{https://www.yomiuri.co.jp/} \copyright~The Yomiuri Shimbun, \url{https://www.rakuten.co.jp/} \copyright~Rakuten Group, Inc., \url{https://www.cosme.net/} \copyright~istyle,inc.
  }
  \label{fig:japanese}
  \Description{A set of four screenshots of various Japanese websites. The screenshots are dense with Japanese text.}
\end{figure*}

An example of aesthetic notions of simplicity differing can be found in Japanese web design. Websites aimed at Japanese audiences are often dense with information and text (Figure~\ref{fig:japanese}). Visitors whose primary experience is of contemporary Euro-American Web design often perceive it as ``cluttered'', and far from simple. But this evaluation applies an inappropriate measure of simplicity. Japanese web design reflects cultural, technical and linguistic differences, encompassing considerations such as the logographic nature of kanji and low-bandwidth mobile data connections \cite{wake_2016, themarketingsamurai_2018, david_2013}. Japan's sociotechnocultural environment identifies simplicity with legibility and information richness \cite{rean_2020, mcgowan_2018}.\footnote{This is not limited to Japan; eye-tracking studies comparing American and Chinese users found radical differences in web page reading patterns, implying that different information layouts are required to achieve the same level of perceived usability for people with different cultural backgrounds \cite{lee2018ai}.} Contrast this with the Euro-American identification of simplicity with whitespace and low information per ``screenful'', typified by websites such as Medium.com. Here the transformation of the Western information aesthetic from print to digital media via the dot-com crisis is clear; Europe had its own long tradition of dense texts (one need only look at the pages of medieval manuscripts, decorated and inscribed in every available space to maximise the precious vellum), which did not survive the transition to the Western Web.






While consumers wish for simple tools, they cannot wish for simple tasks. The challenges and ambitions of work and life are irreducibly complex, and to this we aim to apply simple, intuitive tools. These two forces are in tension, and in the West they have resulted in a ``toolbelt'' style of computing \cite{sumner1997evolution}. This has roots as far back as the UNIX design philosophy that \emph{``A UNIX program should do one thing well, and leave unrelated tasks to other programs''} \cite{pike1984program}. The idea is that each individual piece of software remains small, modular, and robust, but can be composed with others: a workflow that mirrors the epistemic process of pure mathematics. This mathematico-computational view of working is supremely appropriate for mathematicians, but less so for other domains, as acknowledged in Lenat's humility-laden note on \emph{Why AM and EURISKO appear to work} \cite{lenat1984and}. In the late 1990s, Sumner and Stolze characterised this as ushering in the ``toolbelt era'' \cite{sumner1997evolution}, where most complex daily tasks require the coordinated use of several pieces of software, with often jarring frictions and disconnects as the activity transitions between tools. The picture has not changed at the time of writing. Even in praxisware, which often aims to contain as large a portion of common and professional workflows as possible, it is often observed how auxiliary tools are used to finesse, subvert, combine, marshall, and supplement the information workflow in conjunction with the ``primary'' tool \cite{chalhoub2022freedom, larsen2020collaborative}, with ensuing frictions.




The toolbelt philosophy, in turn, puts pressure on knowledge work to accommodate its tendency for modularisation and simplification, thus contributing to a Marxian breakdown of knowledge work \cite{fuchs2014digital}, nudging it in a direction where people only click buttons for small-scope machines to perform simple ``mimeomorphic'' actions \cite{collins1998shape}. A genre of software that has already reached this extreme is social media and recommender systems, where the loss of user agency is well-documented. \v{Z}i\v{z}ek (via Pfaller) describes these as ``interpassive'' \cite{zizek1998interpassive} (by irreverent contrast to inter\emph{active}), giving the user only the slightest illusion of control while scrolling through their Instagram, Twitter, or TikTok feeds, while in practice any given user's experience is driven entirely by the algorithm \cite{hollanek2019non}. The simplification of knowledge work is accompanied by a shift in emphasis from production to consumption. Praxisware concerns itself inherently with production, thus the doctrine of simplicity antagonises its relationship with information and knowledge.




The Chinese software landscape, through a combination of government policy and domestic grit, has developed a compelling alternative to the toolbelt. Known as ``superapps'', platforms such as WeChat encompass, connect, and consolidate wide-ranging functions including communications, payment, and numerous retail and service verticals. As Lee describes \cite{lee2018ai}: \emph{``In effect, WeChat has taken on the functionality of Facebook, iMessage, Uber, Expedia, eVite, Instagram, Skype, PayPal, Grubhub, Amazon, LimeBike, WebMD, and many more. It isn’t a perfect substitute for any one of those apps, but it can perform most of the core functions of each, with frictionless mobile payments already built in. This all marks a stark contrast to the ``app constellation'' model in Silicon Valley in which each app sticks to a strictly prescribed set of functions.''} Moreover, the organisation of these firms acknowledges that information technology is only one participant in a larger world where tasks, goods and services involve flesh and oil: \emph{``American internet companies tend to take a ``light'' approach [...] sharing information, closing knowledge gaps, and connecting people digitally. [...] In China, companies tend to go ``heavy.'' They don’t want to just build the platform—they want to recruit each seller, handle the goods, run the delivery team, supply the scooters, repair those scooters, and control the payment.''} The Chinese example puts a twist on ``appropriate technology'' \cite{schumacher2011small}, a well-intentioned concept often invoked as an epithet meaning ``cheap technology for poor people'', showing that that it is not only the material basis of technology, but also its ideology (such as its aesthetics of simplicity) that may be culturally inappropriate.




Examining the foundations of the doctrine of simplicity also reveals cracks. In particular, the inherent universalism in the ``paradox of the active user'', the idea that users overwhelmingly prefer to learn-by-doing, that they do not wish to invest time in learning, and that ``nobody reads documentation'', is simply false. In her pioneering work with the GenderMag framework \cite{burnett2016gendermag}, Margaret Burnett and her collaborators drew evidence from dozens of studies to show that individuals have a variety of information processing and learning styles. Some indeed prefer to tinker, and learn by doing, eager to seek immediate visible progress. Others are slow and methodical, tend to seek help from others, and look for information and training. GenderMag catalogues these differences and shows that certain styles of learning and information processing are more common in certain demographic segments, with a particular focus on gender. The astonishing and illuminating conclusion is that in privileging certain information processing styles over others, designers introduce gender bias into software. As such, software design in a regime where learning-by-doing is prioritised at the expense of help, documentation, manuals, and training, also introduces biases. There is another chapter to the history outlined in Section~\ref{sec:history}: it was in 1984, amidst the birth of personal computing, that computer science developed its gaping gender disparity \cite{kenney_henn_2014}. Why? Analyses point to a feedback loop seeded by the commercialisation strategy adopted by companies at the time: to sell computers to children as toys. But toy advertising is notoriously gendered, and in this precise moment, the industry ``picked a side'' and decided that computing was for boys \cite{misa2011gender}, setting off a cascade of downstream effects where women entered computing majors with much less prior experience than their peers identifying as men. Thus each time a new, more democratic (and therefore lucrative) generation of interface arises, commercial pressures intercede to privilege retail success over user experience, excluding whoever need be. And so the market pressure for simplicity exerts normative and exclusionary effects on praxisware and its communities.


\section{Towards a doctrine of negotiated complexity}

In the genesis of the minimal learning paradigm, researchers had already identified that the tension between investing in learning materials, versus investing in simplicity by design, would be a problem for praxisware. In 1995, Sellen and Nicol concluded \cite{sellen1995building}: \emph{``The distinction between on-line help [i.e., interactive documentation] and the user interface is not necessarily clear-cut. [...] The aim should be to provide simple, self-explanatory interfaces. Although the ideal is commendable, reaching it seems unrealistic at the moment, especially considering the imbalance between the power and flexibility that much contemporary software affords and the state of the art in intelligent interfaces.''}



Nearly thirty years later, it is clear that just advancing the ``state of the art in intelligent interfaces'' is unlikely to ever be sufficient for tackling the ``power and flexibility'' of praxisware. One design response attempting to bridge the obvious dissonance has been the rise of ``in-app teaching'', such as call-outs and dialogs announcing and explaining new features. Yet these still suffer from the cultural hangover of minimal learning, and often are heavily constrained in the amount of information they present to the user. Competing for attention in a digital landscape overloaded with notifications \cite{Toffler1970-kq,biskjaer2016taking}, these simple interventions are often ignored entirely.

A deeper limitation of in-app teaching is the premise that users need to learn just-in-time, when they are at their computers to perform a specific task. This is a powerful form of learning because it can take advantage of the intrinsic motivation and context to complete a particular task, but can be inhibited by the same need, if the opportunity to learn is seen as a distraction. The success or failure of such interventions hinges on accurately modelling the user's current knowledge as well as their intent, both of which are exquisitely challenging.


The alternative is for the user to learn ahead of time, perhaps with the background context of a set of certain domain-relevant tasks, but where the tasks are not ends in themselves. This shifts learning from being task-oriented to being capability- or vision-oriented. Knowing what can be done, an \emph{awareness of the possible}, empowers users in entirely different ways. As Rittel and Webber note \cite{rittel1973dilemmas}, conceiving a problem in a certain way prefigures its solution. It shifts the nature of the task itself. A spreadsheet user who knows \emph{a priori} how to write formulas will approach the same task differently from one who does not; starting with planning the data that must go in the spreadsheet and structuring it appropriately \cite{chalhoub2022freedom}. By the time the spreadsheet is open and ready, the ``task'' has already taken on a vastly different nature for the two users. And on a moment-by-moment basis, the patchwork of methods and available options explored by each user is also radically different, shaped by their differing awareness of possibilities adjacent to the current state of their spreadsheet \cite{pandita2018no}. Ahead of time, task-independent learning complements the just-in-time approach to facilitate these shifts in awareness and task conceptualisation.



The fact that \emph{some} users are willing to invest time and effort into learning is already documented, but assumed to be rare and not reflective of mainstream behaviour, and relying on this willingness is seen as a risk to commercial success. However, recent trends on the social media platforms TikTok and Instagram reveal that the desire for ahead of time learning might be far more widespread than previously thought. I am referring to the meteoric rise of praxisware influencers, who present short tutorials on complex software such as Excel or Photoshop, often accompanied by jaunty tunes and dancing (as per the TikTok vernacular style), some of whom have followers numbering in \emph{millions} \cite{press-reynolds_2021}. Unlike a platform like YouTube, where tutorials can actively be searched for and watched on demand, the engagement patterns of TikTok and Instagram are those of passive, serendipitous consumption. Why would millions of people follow a TikTok influencer who teaches Excel? It is not because they expect to stumble upon a video that will help them with a task they're currently working on. It is because in moments of passive consumption, on the sofa, on the commute, they expect to benefit from ahead of time learning, and enhancing their awareness of the possible. These millions of subscribers are strong evidence of the mainstream desire, and acceptance, of learning investment for praxisware.


We must re-evaluate our attitude to tutorials, learning materials, and the user learning journey, and reconsider their position as a key part of interface design. We must also re-evaluate our attitude to the ideals of simplicity in software use. In this paper I have proposed that ``easy to use'' is an inappropriate criterion for evaluating complex, feature-rich praxisware. This is a call for \emph{negotiated complexity}, complexity in software which arises as the result of optimisation for intensive, long-term use, in consultation with communities of practitioners, shielded from commercial pressure to design for first impressions, and with the unapologetic assumption that users will invest months, or years, in ahead of time learning and deliberate practice. Complexity is negotiated as the acceptable trade-off between time invested in learning a tool and the resultant power it gives the practitioner. Complexity is negotiated between designers and practitioners over long-term use, not fixed, and evolves with the needs of the community. The agenda for negotiated complexity in praxisware has among its implications for research and design the following:
\begin{itemize}
    \item Rather than usability as a universal property of a user interface with respect to some notion of what a monolithic user might find `intuitive' and `natural', we define usability as a function of an interface paired with well-designed learning and instructional materials.
    \item We consider investment in learning and instruction as a first-class citizen that participates with and complements the design of the system.
    \item We reward longitudinal research that engages seriously with practitioner communities, and accommodate it within the academic cycle, perhaps by developing new methods for evaluating the contribution of an ongoing longitudinal study, or new tracks that prioritise such work.
    \item We begin to shift customer culture, when appropriate, to redress the adverse commercial incentives for simplicity, and to save ourselves from the prisoner's dilemma. While this may seem challenging, loci of deliberate learning such as TikTok seem like a good place to start.
\end{itemize}

In this paper, I have described the doctrine of simplicity, which has come to be a dominant ideal in user interface research and design. By tracing key events in the history of computing and design research, I have attempted to explain how this ideal has taken hold, at least in the West. Such ideals create a self-sustaining market pressure that often countermands the design of complex, feature-rich software. Such software, termed praxisware, has several properties which cause it to be vulnerable to the doctrine of simplicity. For praxisware, a better ideal to aim for is negotiated complexity, considering user learning as part and parcel of user interface design.

\begin{acks}
Thanks to Roger Yin for conversations about the prisoner's dilemma, to Andy Gordon and Ian Drosos for their thoughts on drafts of the paper, and to my reviewers for their helpful critiques.
\end{acks}

\bibliographystyle{ACM-Reference-Format}
\bibliography{sample-base,references}










\end{document}